\documentclass[%
 reprint,
 amsmath,amssymb,
 aps,
]{revtex4-2}

\usepackage{graphicx}
\usepackage{dcolumn}
\usepackage{bm}
\usepackage{hyperref}

\usepackage{slashed}
\usepackage{float}
\begin{document}


\title{Vortical effects and the critical end point in the Linear Sigma Model coupled to quark}

\author{L. A. Hern\'andez}
 \affiliation{Departamento de F\'isica, Universidad Aut\'onoma Metropolitana-Iztapalapa, Avenida San Rafael Atlixco 186, Ciudad de México 09340, Mexico.}
\author{R. Zamora}%
\affiliation{Instituto de Ciencias B\'asicas, Universidad Diego Portales, Casilla 298-V, Santiago, Chile.\\}%
\affiliation{Facultad de Ingenier\'ia, Arquitectura y Diseño, Universidad San Sebasti\'an, Santiago, Chile.}%


\begin{abstract}
In this paper, we study the effects of vorticity on the QCD phase transition using the Linear Sigma Model coupled to quarks. By going beyond the mean-field approximation and incorporating screening effects via ring diagrams, we explore the chiral symmetry restoration in extreme conditions, such as high temperatures, high densities, and large angular velocities. Our analysis reveals how the critical temperature decreases as the angular velocity increases, suggesting that vorticity catalyzes the symmetry restoration. Additionally, we observe a shift in the Critical End Point (CEP) in the effective QCD phase diagram, where higher angular velocities move the CEP to lower quark chemical potentials and higher temperatures. Moreover, we analyze the baryon number fluctuations through the normalized fourth moment  $\kappa \sigma^2 = c_4/c_2$ as a function of the collision energy in heavy-ion reactions $\sqrt{s_{NN}}$, which serves as a key observable to identify the CEP. Our study reveals that for high collision energies,  $\kappa \sigma^2$  remains nearly constant; however, as the system approaches the CEP, the ratio increases sharply, indicating the proximity of the critical region. This rise is influenced by the presence of vorticity, which causes the CEP to shift to higher collision energies. These findings provide insight into the role of vorticity in heavy-ion collisions
\end{abstract}

\maketitle


\section{Introduction \label{sec1}}

In everyday life, it is relatively easy to observe systems with finite vorticity. Atmospheric phenomena such as cyclones or hurricanes are common examples. Also, we can observe these kind of phenomena at different scales like the rotation of galaxies, the Great Red Spot on Jupiter or electrons flowing in vortices. However,  the system that achieves the highest angular velocity occurs in relativistic heavy-ion collisions, when the collision is non-central, we observe an angular velocity $\Omega \simeq (9 \pm 1) \times 10^{21} s^{-1}$ which is equal to $\Omega \simeq 7$ MeV~\cite{STAR:2017ckg}. The magnitude of this angular velocity is enough to think in possible effects on the dynamics of the reaction, where a vortical fluid can be created. One of the most relevant observables where the effect of this vorticity has been studied is the global particle polarization. This possibility has prompted the search for global hadron polarization, the most notable study being that of hyperons $\Lambda$ and $\bar{\Lambda}$ polarization~\cite{Xie:2015xpa,Xie:2016fjj,Becattini:2016gvu,Sorin:2016smp,Li:2017slc,Sun:2017xhx,Han:2017hdi,Xia:2018tes,Baznat:2017jfj,Karpenko:2018erl,Suvarieva:2018wmh,Kolomeitsev:2018svb,Xie:2019jun,Ayala:2020soy,Ayala:2021xrn}. The idea of high vorticity in relativistic heavy ion collisions suggests the possibility of other possible effects. One of them is to analyze a possible modification of the phase transition curve in the QCD phase diagram, as a consequence of varying the angular velocity of the reaction. Elements that support this idea comes from the result obtained from different simulations of the angular velocity profile parametrized by the collision's proper time, where it has a small decrease at the time the quark-hadron phase transition is generated, especially at low energies at the center of mass of the collision~\cite{Deng:2016gyh,Deng:2020ygd}. Thus, we explore the possibility of observing a shift in the position of the \textit{critical end point} (CEP) in the QCD phase diagram as a result of the vorticity in the medium. There are already work which are studied the QCD phase transition in the presence of $\Omega$~\cite{Jiang:2016wvv,Wang:2018sur,Chernodub:2016kxh,Chen:2023cjt,Sun:2023kuu,Fujimoto:2021xix,Chen:2020ath,Braga:2022yfe,Wang:2024szr,Yamamoto:2013zwa,Braguta:2020biu,Braguta:2021jgn,Yang:2023vsw,Sun:2024anu,Cao:2023olg,Jiang:2023zzu,Gaspar:2023nqk,TabatabaeeMehr:2023tpt,Chernodub:2020qah,Chernodub:2022veq,Braguta:2023iyx}. However, it is true that our knowledge of the QCD phase diagram at large values of baryonic density is poor, even at zero angular velocity and just exploring the $T-\mu_B$ plane. Therefore works where the phase transition is studied and includes analysis considering a rich baryonic matter are completely suitable. This idea is reinforce knowing that new facilities are under construction such as NICA~\cite{MPD:2022qhn} and FAIR~\cite{Agarwal:2022ydl}, and taking into account the multi-step program of RHIC, the Beam Energy Scan~\cite{WPBES}; all of them with the goal to explore deep in the rich baryonic matter region.

In this work, we explore the consequences that emerge from the inclusion of vortical effects in the dynamics of a strongly interacting matter in extreme conditions, such as high temperatures and densities. Our main objectives are first to obtain the transition curves for given angular velocity values and in parallel to identify the type of transition that is occurring, in order to observe possible changes in the CEP location when the angular velocity changes. One of the key elements to highlight in this work is the inclusion of the screening effects, we are going to go beyond the mean field theory, it allows to explore with a more realistic analysis the QCD phase transition. Also, we are including vortical effects in both the scalar field propagator~\cite{Gaspar:2023nqk} and the fermion field propagator~\cite{Ayala:2021osy}. Another important element is that in order to manage analytic expressions, we decide to work in the high temperature limit, it allows us tracking the features behind the results that we obtain.

The work is organized as follows. In Section \ref{sec2}, we introduce the Linear Sigma Model coupled to quarks. Then, in Section \ref{sec3}, we compute the effective potential up to ring diagrams contribution at finite temperature, quark chemical potential, and angular velocity, and we show the corresponding effective QCD phase diagram. In Section \ref{sec4}, we formulate the way the baryon number fluctuations can be described in terms of the probability distribution associated to the order parameter near the transition line and present the result of the analysis for the ratio of the cumulants $c_4/c_2$ as a function of $\sqrt{s_{NN}}$ in order to locate the CEP in term of the energy at the center of mass of the collision. Finally, in Section \ref{sec5} and \ref{sec6}, we will present the discussion of our results and conclusions, respectively.

\section{Linear Sigma Model coupled to quarks \label{sec2}}
The Linear Sigma Model coupled to quarks (LSMq) is an effective theory which is useful to emulate the low energy regime of Quantum Chromodynamics. The degrees of freedom of this model is a mixture of scalar and pseudo-scalar mesons and the two lightest quark flavors. The main feature of the LSMq is that it can exhibit a symmetry spontaneously broken. The Lagrangian for this model is the following
\begin{align}
    \mathcal{L}&=\frac{1}{2}(\partial_\mu \sigma)^2  + \frac{1}{2}(\partial_\mu \vec{\pi})^2 + \frac{a^2}{2} (\sigma^2 + \vec{\pi}^2)\nonumber\\
   &- \frac{\lambda}{4} (\sigma^2 + \vec{\pi}^2)^2 + i \bar{\psi} \gamma^\mu \partial_\mu\psi -g\bar{\psi} (\sigma + i \gamma_5 \vec{\tau} \cdot \vec{\pi} )\psi ,
\label{lagrangian}
\end{align}
where $\psi$ is an $SU(2)$ isospin doublet of quarks, $\sigma$ is an isospin singlet and $\vec{\pi}=(\pi_1, \pi_2, \pi_3 )$ is an isospin triplet, corresponding to the sigma meson and three neutral pions, respectively. In Eq.~(\ref{lagrangian}), $\tau$ are the Pauli matrices. Also, two different couplings appear, $\lambda$ and $g$, the boson self-coupling and the fermion-boson coupling, respectively. The squared mass parameter is $a^2$. In this work, we take $a^2,\lambda,g >0$. Additionally, in nature, we observe two charged pions and one neutral pion. Therefore, we implement the following transformation
\begin{equation}
 \pi_\pm=\frac{1}{\sqrt{2}}(\pi_1\pm i\pi_2).
\end{equation}
After this transformation and letting the $\sigma$ field to develop a vacuum expectation value $v$, namely
\begin{equation} 
\sigma \rightarrow \sigma + v,
\label{shift}
\end{equation} 
in order to allow for a spontaneous symmetry breaking. We rewrite the Lagrangian in Eq.~(\ref{lagrangian}) as follows
\begin{align}
    \mathcal{L}&=\frac{1}{2}\partial_{\mu}\sigma \partial^{\mu}\sigma+\frac{1}{2}\partial_{\mu}\pi_{0}\partial^{\mu}\pi_{0}+\partial_{\mu}\pi_{-}\partial^{\mu}\pi_{+}\nonumber\\
    &-\frac{1}{2}m_{\sigma}^{2}\sigma^{2}-\frac{1}{2}m_{0}^{2}\pi_{0}^{2}-m_{0}^{2}\pi_{-}\pi_{+}+i\bar{\psi}\slashed{\partial}\psi\nonumber\\
    &-m_{f}\bar{\psi}\psi+\frac{a^2}{2}v^2-\frac{\lambda}{4}v^4+\mathcal{L}_{int},
    \label{linearsigmamodelSSB}
\end{align}
where the interaction Lagrangian is defined as
\begin{align}
    \mathcal{L}_{int}&=-\frac{\lambda}{4}\sigma^{4}-\lambda v\sigma^{3}-\lambda v^{3}\sigma-\lambda\sigma^{2}\pi_{-}\pi_{+} -2\lambda v \sigma\pi_{-}\pi_{+} \nonumber \\
    &-\frac{\lambda}{2}\sigma^{2}\pi_{0}^{2}-\lambda v\sigma \pi_{0}^{2}-\lambda \pi_{-}^{2}\pi_{+}^{2}-\lambda\pi_{-}\pi_{+}\pi_{0}^{2}-\frac{\lambda}{4}\pi_{0}^{4} \nonumber \\ 
    &+a^{2}v\sigma -g\bar{\psi}\psi\sigma-ig\gamma^{5}\bar{\psi}\left(\tau_{+}\pi_{+}+\tau_{-}\pi_{-}+\tau_{3}\pi_{0}\right)\psi.
    \label{interactinglagrangian}
\end{align}
As can be seen from Eqs.~(\ref{linearsigmamodelSSB}) and~(\ref{interactinglagrangian}), there are new terms that depend on $v$ and all fields develop dynamical masses, 
\begin{align}
     m_{\sigma}^{2}&=3\lambda v^2-a^2, \nonumber \\
     m_{\pi}^{2}&=\lambda v^2-a^2, \nonumber \\ 
     m_{f}&=gv.
\label{masses}
\end{align}
Since the dynamical masses depend on the vacuum expectation value, $v$, and the latter is the order parameter associated to the symmetry which is spontaneously broken. Then, we identify that it is the chiral symmetry. On the other hand, from Eq.~(\ref{linearsigmamodelSSB}), we also notice that we have a term that describe the shape of the potential along the direction where the symmetry was broken, it is call the three level or the classical potential which has the expression
\begin{equation}
    V^{\text{tree}}(v)=-\frac{a^2}{2}v^2+\frac{\lambda}{4}v^4,
    \label{treelevel}
\end{equation}
whose minimum is found at
\begin{equation}
    v_0=\sqrt{\frac{a^2}{\lambda}}.
\end{equation}
Since $v_0\neq 0$, we notice that chiral symmetry is spontaneously broken. To determine the conditions for chiral symmetry restoration as a function of $T$, $\Omega$ and $\mu_q$, we study the behavior of the effective potential which, for this work, includes the classical potential or tree-level contribution, the one-loop correction both for bosons and fermions and the ring diagrams contribution, which accounts for the plasma screening effects. In the next section we compute each of these contribution, once the effective potential is computed, we proceed to find the transition curves in the phase diagram temperature vs quark chemical potential, varying the angular velocity.

\section{Effective QCD phase diagram \label{sec3}}

To study chiral symmetry restoration, we begin by computing the effective potential, including contributions up to the ring diagrams. It means we go beyond mean field approximation, considering the plasma screening effects. The effective potential structure is described by
\begin{equation}
    V^{\text{eff}}=V^{\text{tree}}+V^1_b+V^1_f+V^{\text{rings}}. 
    \label{Veffstructure}
\end{equation}

In this work, we present analytic expressions for each contribution in Eq.~(\ref{Veffstructure}). It implies that some approximation are necessary. We compute the effective potential in the high temperature approximation, working within the Matsubara formalism. The starting expressions that we are going to use are: for the 1-loop boson contribution, we have
\begin{equation}
    V_b^1=T \sum_n \int \frac{d^3k}{(2\pi)^3} \ln  D^\Omega (\omega_n,\vec{k})^{1/2},
    \label{V1binitial}
\end{equation}
with
\begin{equation}
    D^\Omega (\omega_n,\vec{k})=-\frac{1}{(\omega_n-i\Omega)^2+k_\perp^2+k_z^2+m_b^2}.
    \label{propBvorticity}
\end{equation}
For the 1-loop fermion contribution, we have
\begin{equation}
    V_f^1=-T\sum_n \int \frac{d^3k}{(2\pi)^3} \text{Tr}\big[ \ln S^\Omega(\tilde{\omega}_n,\mu_q,\vec{k})^{-1} \big],
    \label{V1finitial}
\end{equation}
with
\begin{widetext}
\begin{align}
    S^\Omega(\tilde{\omega}_n,\mu_q,\vec{k})=&-\frac{(i\tilde{\omega}_n+\mu_q+\Omega/2-k_z+ik_\perp)(\gamma_0+\gamma_3)+m_f(1+\gamma_5)}{(\tilde{\omega}_n-i\mu_q-i\Omega/2)^2+\vec{k}^2+m^2}\mathcal{O}^+ \nonumber \\
    &-\frac{(i\tilde{\omega}_n+\mu_q-\Omega/2+k_z-ik_\perp)(\gamma_0-\gamma_3)+m_f(1+\gamma_5)}{(\tilde{\omega}_n-i\mu_q+i\Omega/2)^2+\bar{k}^2+m^2}\mathcal{O}^-.
    \label{finalfermionpropagator}
\end{align} 
\end{widetext}
For the ring diagrams contribution, we have
\begin{equation}
    V^{\text{rings}}=\frac{T}{2}\sum_n \int \frac{d^3k}{(2\pi)^3} \ln \big (1+\Pi \  D^\Omega(\omega_n,\vec{k}) \big),
    \label{Vringsinitial}
\end{equation}
with $\Pi$ the boson's self-energy. In Eqs.~(\ref{V1binitial}),~(\ref{V1finitial}) and~(\ref{Vringsinitial}); $\omega_n$ and $\tilde{\omega}_n$ are the Matsubara frequencies for bosons and fermions, respectively. The boson propagator in Eq.~(\ref{propBvorticity}) is taken from Ref.~\cite{Gaspar:2023nqk} and the fermion propagator in Eq.~(\ref{finalfermionpropagator}) was reported in Ref.~\cite{Ayala:2021osy}. In both cases, the propagators are immersed within a rigidly rotating environment with cylindrical geometry, where $\Omega$ is the angular velocity. For the fermionic contribution, we also have extra ingredients, they are the quark chemical potential $\mu_q$ and the projectors $\mathcal{O}^\pm$ which are defined as
\begin{equation}
    \mathcal{O}^\pm=\frac{1}{2}(1\pm i\gamma^1\gamma^2).
\end{equation}

The computation of the effective potential is performed under the approximation that the temperature is an energy scale satisfying the condition $T\gg m_b$, with $b=\sigma, \pi, q$. The first contribution to calculate is the 1-loop boson term, Eq.~(\ref{V1binitial}), which can be written as
\begin{equation}
    V_b^1=\frac{T}{2}\int \frac{d^3 k}{(2\pi)^3}\int dm_b^2 \sum_n \frac{1}{(\omega_n-i\Omega)^2+\vec{k}^2+m_b^2}.
\end{equation}
As a first step, we sum over the Matsubara frequencies, getting
\begin{equation}
    V_b^1=\frac{1}{2}\int \frac{d^3 k}{(2\pi)^3}\int dm_b^2 \frac{1}{2 E}[1+n_b(E-\Omega)+n_b(E+\Omega)],
    \label{V1bdistributions}
\end{equation}
where $E=\sqrt{\vec{k}^2+m_b^2}$ and $n_b(E\pm \Omega)$ are the corresponding Bose-Einstein distributions. From Eq.~(\ref{V1bdistributions}), we notice that there are two kind of terms, the first one inside of the brackets which is T independent and the last to terms inside the brackets which are T dependent. We called the former vacuum piece and the latter matter piece. Therefore, we can write the following
\begin{equation}
    V_b^1=V^1_{b,\text{vac}}+V^1_{b,\text{mat}}.
\end{equation}
The vacuum term, $V^1_{b,\text{vac}}$, has a UV divergence, which we handle accordingly. We use dimensional regularization to isolate the divergence, renormalize the term, and implement the $\overline{MS}$ subtraction scheme. After completing this process, we obtain
\begin{equation}
    V^1_{b,\text{vac}}=\frac{m_b^4}{64\pi^2}\bigg[ \ln \bigg( \frac{\mu^2}{m_b^2} \bigg)+\frac{3}{2} \bigg],
    \label{V1bvacfinal}
\end{equation}
where $\mu$ is the renormalization scale. The matter piece is computed following the prescription of $T\gg m_b$ and the final expression for each boson flavor is given by~\cite{Gaspar:2023nqk}.
\begin{align}
   V^1_{b,\text{mat}}&=-\frac{\pi^2 T^2}{90}+\frac{T^2 (m_b^2-2\Omega^2)}{24}-\frac{T}{12\pi}(m_b^2-\Omega^2)^{3/2}\nonumber \\
   &+\frac{\Omega^4}{48\pi^2}-\frac{\Omega^2m_b^2}{16\pi^2}-\frac{m_b^4}{64\pi^2}\bigg [ \ln \bigg ( \frac{m_b^2}{16\pi^2 T^2} \bigg ) +2\gamma_E-\frac{3}{2} \bigg], 
   \label{V1bmatfinal}
\end{align}
where $\gamma_E$ denoting the Euler-Mascheroni constant. We put together both vacuum and matter pieces and the 1-loop boson term becomes
\begin{align}
    V^1_b&=-\frac{\pi^2 T^4}{90}+\frac{T^2 (m_b^2-2\Omega^2)}{24}-\frac{T}{12\pi}(m_b^2-\Omega^2)^{3/2}\nonumber \\
   &+\frac{\Omega^4}{48\pi^2}-\frac{\Omega^2m_b^2}{16\pi^2}-\frac{m_b^4}{64\pi^2}\bigg [ \ln \bigg ( \frac{\mu^2}{16\pi^2 T^2} \bigg )+2\gamma_E \bigg].
   \label{V1bfinal}
\end{align}

Now, we proceed to compute the 1-loop fermion contribution, where the way is completely analogous to the boson case. We rewrite the Eq.~(\ref{V1finitial}) as follows
\begin{align}
    V_{\text{f}}^1=-T \sum_n \int dm_f^2 \int \frac{d^3k}{(2\pi)^3} &\Big [ \frac{1}{(\omega_n-i\mu_1)^2+\vec{k}^2+m_f^2} \nonumber \\
    &+ \frac{1}{(\omega_n-i\mu_2)^2+\vec{k}^2+m_f^2} \Big], 
\end{align}
with
\begin{equation}
    \mu_1= \mu_q+\frac{\Omega}{2}, \ \ \ \ \ \ \ \mu_2= \mu_q-\frac{\Omega}{2}.
\end{equation}
We perform the sum over the Matsubara frequencies and $V_{\text{f}}^1$ becomes
\begin{align}
    V_{\text{f}}^1&=-\int dm_f^2 \int \frac{d^3k}{(2\pi)^3}\frac{1}{2E}\nonumber \\
    &\times \bigg[ \Big(1-n_f(E+\mu_1)-n_f(E-\mu_1) \Big)\nonumber \\
    &+\Big(1-n_f(E+\mu_2)-n_f(E-\mu_2) \Big)\bigg],
    \label{V1fdistributions}
\end{align}
where $E=\sqrt{\vec{k}^2+m_f^2}$ and $n_f(E\pm \mu_{1,2})$ are the corresponding Fermi-Dirac distributions. Equation~(\ref{V1fdistributions}) can also be split in the vacuum and matter terms. The vacuum contribution is
\begin{align}
    V^1_{\text{f,vac}}&=-\int dm_f^2 \int \frac{d^3k}{(2\pi)^3}\frac{1}{2\sqrt{\vec{k}^2+m_b^2}}\nonumber \\
    &=\frac{m_f^4}{16\pi^2}\bigg[ \ln\Big( \frac{\mu^2}{m_f^2} \Big)+\frac{3}{4} \bigg],
\end{align}
where one more time for the vacuum piece we use dimensional regularization in order to isolate the divergence, renormalize this term and implement the subtraction scheme $\overline{MS}$, with $\mu$ the renormalization scale. The matter contribution is computed in the high $T$ approximation, which takes into account the relation $T\gg m_f$. Detailed calculation can be found in Appendix~\ref{Appe1}. Here, we write the final expression
\begin{align}
    V^1_{\text{f,mat}}=&\frac{m_f^4}{16\pi^2}\bigg [ \ln \Big(\frac{m_f^2}{\pi^2 T^2} \Big) + 2\gamma_E -\frac{3}{2} \bigg]-\frac{7 T^4 \pi^2}{180}\nonumber \\
&-\frac{T^2}{12}\Bigg( \Big( \mu_q+\frac{\Omega}{2} \Big)^2+\Big( \mu_q-\frac{\Omega}{2} \Big)^2 \Bigg)\nonumber \\
&-\frac{T^2m_f^2}{4\pi^2}\Bigg ( \text{Li}_2\left(-e^{\frac{\mu +\frac{\Omega }{2}}{T}}\right)+\text{Li}_2\left(-e^{\frac{\mu -\frac{\Omega }{2}}{T}}\right)\nonumber \\
&+\text{Li}_2\left(-e^{-\frac{\mu +\frac{\Omega }{2}}{T}}\right)+\text{Li}_2\left(-e^{-\frac{\mu -\frac{\Omega }{2}}{T}}\right) \Bigg ) \nonumber \\
&-\frac{\left(\mu +\frac{\Omega }{2}\right)^4+\left(\mu -\frac{\Omega }{2}\right)^4}{24 \pi ^2}. \label{materiaf}]
\end{align}

In order to go beyond mean field approximation, we include the ring diagrams contribution, it implements the screening effects in the medium. Since, we work with the high T limit, we only consider the dominant term in Eq.~(\ref{Vringsinitial}), it is the zero Matsubara mode. Hence, we have
\begin{equation}
    V^{\text{ring}}=\frac{T}{2} \int \frac{d^3k}{(2\pi)^3} \ln (1+ \Pi D^\Omega(\omega_0,\vec{k})),
    \label{VringsHighT}
\end{equation}
with $\Pi$ the boson’s self-energy, which in the high temperature limit, is given by (the computation details are reported in Appendix \ref{appe2})
\begin{eqnarray}
\Pi&=& \frac{\lambda T^2}{2} -\frac{N_f N_c T^2 g^2}{2 \pi^2} \Bigg ( \text{Li}_2\left(-e^{\frac{\mu +\frac{\Omega }{2}}{T}}\right)+\text{Li}_2\left(-e^{\frac{\mu -\frac{\Omega }{2}}{T}}\right)\nonumber \\
&+&\text{Li}_2\left(-e^{-\frac{\mu +\frac{\Omega }{2}}{T}}\right)+\text{Li}_2\left(-e^{-\frac{\mu -\frac{\Omega }{2}}{T}}\right) \Bigg ).
\label{fullexpressionselfenergy}
\end{eqnarray}
In order to get the final expression of Eq.~(\ref{VringsHighT}), we perform the angular integrals and it takes the form
\begin{align}
    V^{\text{ring}}=&\frac{T}{4\pi^2} \int dk \ k^2\big[ \ln(\Omega^2+\vec{k}^2+m_b^2+\Pi)\nonumber \\
    &-\ln(\Omega^2+\vec{k}^2+m_b^2)\big],
\end{align}
performing the integral over the momentum, we finally get
\begin{equation}
    V^{\text{ring}}=-\frac{T}{12\pi}(m_b^2-\Omega^2+\Pi)^{3/2}+\frac{T}{12\pi}(m_b^2-\Omega^2)^{3/2}.
\end{equation}

We have all the contributions up to ring diagrams at hand. Thus, we proceed to join all the calculated pieces and write the full expression, taking into account all the degrees of freedom in the LSMq, of the effective potential in the high temperature limit
\begin{align}
    V^{\text{eff}}&=-\frac{a^2}{2}v^2+\frac{\lambda}{4}v^4+\sum_{b=\sigma,\vec{\pi}} \Bigg \{- \frac{m_b^4 }{64 \pi ^2} \bigg[\ln \left(\frac{\mu ^2}{16 \pi^2T^2}\right)\nonumber \\
    &+2\gamma_E\bigg]-\frac{\pi ^2 T^4}{90}+\frac{T^2}{24}\left(m_b^2-2 \Omega ^2\right)\nonumber \\
    &-\frac{T \left(\Pi+m_b^2-\Omega ^2\right)^{3/2}}{12 \pi }-\frac{\Omega ^2 }{48 \pi ^2}\left(3 m_b^2-\Omega ^2\right)\nonumber \Bigg \}\nonumber \\
    &+N_f N_c \Bigg \{ \frac{m_f^4}{16\pi^2} \bigg[\ln \left(\frac{\mu^2}{\pi^2T^2}\right) +2\gamma_E-\frac{3}{4}\bigg] -\frac{7 T^4 \pi^2}{180}\nonumber \\&-\frac{T^2}{12}\Bigg( \Big( \mu_q+\frac{\Omega}{2} \Big)^2+\Big( \mu_q-\frac{\Omega}{2} \Big)^2 \Bigg)\nonumber \\ 
    &-\frac{T^2m_f^2}{4\pi^2}\Bigg ( \text{Li}_2\left(-e^{\frac{\mu +\frac{\Omega }{2}}{T}}\right)+\text{Li}_2\left(-e^{\frac{\mu -\frac{\Omega }{2}}{T}}\right)\nonumber \\
&+\text{Li}_2\left(-e^{-\frac{\mu +\frac{\Omega }{2}}{T}}\right)+\text{Li}_2\left(-e^{-\frac{\mu -\frac{\Omega }{2}}{T}}\right) \Bigg ) \nonumber \\
&-\frac{\left(\mu +\frac{\Omega }{2}\right)^4+\left(\mu -\frac{\Omega }{2}\right)^4}{24 \pi ^2} \Bigg \}.
\label{fullpotential}
\end{align}
\begin{figure}[t]
    \centering
    \includegraphics[scale=0.58]{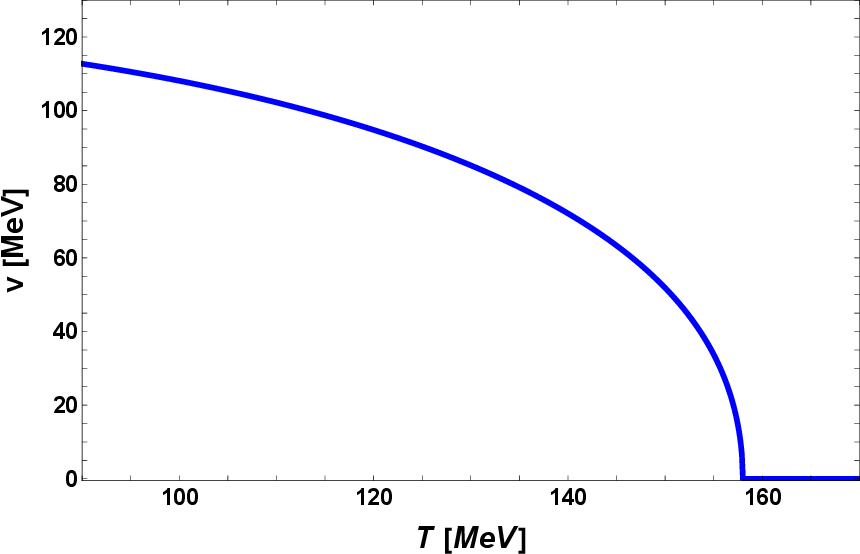}
    \caption{Vacuum expectation value (the order parameter of the theory) as a function of the temperature, for $\mu_q=0$ MeV, $\Omega=0$ MeV, $\lambda=1.4$, $g=0.88$ and $\mu=500$ MeV. We can notice a continuous change of $v$ as a function of $T$. However, the first derivative has a discontinuity when $v\rightarrow 0$. Therefore, this plot shows a second-order phase transition.}
    \label{fig1}
\end{figure}

\begin{figure}[t]
    \centering
    \includegraphics[scale=0.58]{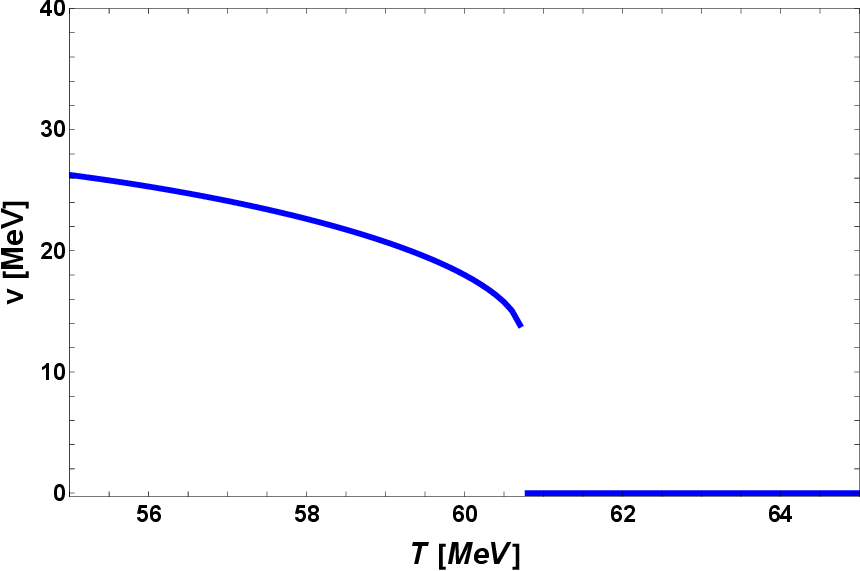}
    \caption{Vacuum expectation value (the order parameter of the theory) as a function of the temperature, for $\mu_q=273$ MeV, $\Omega=0$ MeV, $\lambda=1.4$, $g=0.88$ and $\mu=500$ MeV. We can continuous change of $v$ up to it takes a cero value. This behavior exhibits a discontinuity in the $v$ as a function of $T$. Therefore, this plot shows a first-order phase transition.}
    \label{fig2}
\end{figure}

We use the Eq.~(\ref{fullpotential}) to identify the phase transition associated with the restoration of the chiral symmetry. The way we do the analysis is tracking the evolution of the vacuum expectation value $v$, which is the order parameter of the theory as a function of the thermodynamics variables. We start fixing the value of the angular velocity $\Omega$. Next, we vary the quark chemical potential $\mu_q$ and, for each value, we determine the critical temperature $T_c$. As a result, we obtain a transition curve within the phase diagram in the $T-\mu_q$ plane. The relevance of following the vacuum expectation value is not only to identify when the phase transition happens but also to know the kind of phase transition. In Figs.~\ref{fig1} and~\ref{fig2}, we plot the order parameter as a function of the temperature, given the values of $\mu_q$ and $\Omega$. We observe that in Fig.~\ref{fig1}, $\mu_q=0$ MeV (low baryonic density) the behavior of $v$ is continuous but its derivative shows a discontinuity at the critical temperature (the temperature when $v$ becomes zero). Hence, we say that a second-order phase transition occurs. For Fig.~\ref{fig2}, we depict the same kind of plot. However, in this case, with $\mu_q=273$ MeV (high baryonic density), we have a first-order phase transition, since we can observe the behavior of $v$ is discontinuous itself, and this discontinuity happens at the critical temperature.

Once we are able to identify the phase transitions lines and the kind of these, we proceed to generate an effective QCD phase diagram. However, we should notice that the LSMq contains three independent parameters, namely, the Lagrangian squared mass parameter $a^2$ and the boson and fermion-boson couplings $\lambda$ and $g$. For a complete description of the phase diagram, these parameters need to be fixed using conditions suitable for finite $T$ and $\mu_q$, and not from vacuum conditions. The way that we decide to fix the free parameters consists in two parts. The first one is to use the information given by LQCD which tells us that the pseudo-critical temperature at zero baryonic chemical potential es $T_c\simeq 158$ MeV and provides the curvature parameters $\kappa_2$ and $\kappa_4$~\cite{Borsanyi:2020fev}. We can compute the values of $\lambda$, $g$ and $a$ that best describe the transition curve near $\mu_q\approx 0$. Since, at low values of $\mu_q$ the behavior of the vacuum expectation value does not show any discontinuity and the effective potential for every temperature exhibits only one minimum. Then, at the phase transition we observe a flat potential at $v=0$. All of these features are satisfied if we ask that the square of the boson thermal mass, $\Pi+m_b^2$, vanishes for $v=0$ and $T=T_c$, where the expression for the self-energy was shown in Eq.~(\ref{fullexpressionselfenergy}). The solution obtained from the conditions mentioned is not unique, thus we proceed to show a couple of phase diagrams with different set of parameters in order to show the stability of the results.
\begin{figure}[t]
    \centering
    \includegraphics[scale=0.58]{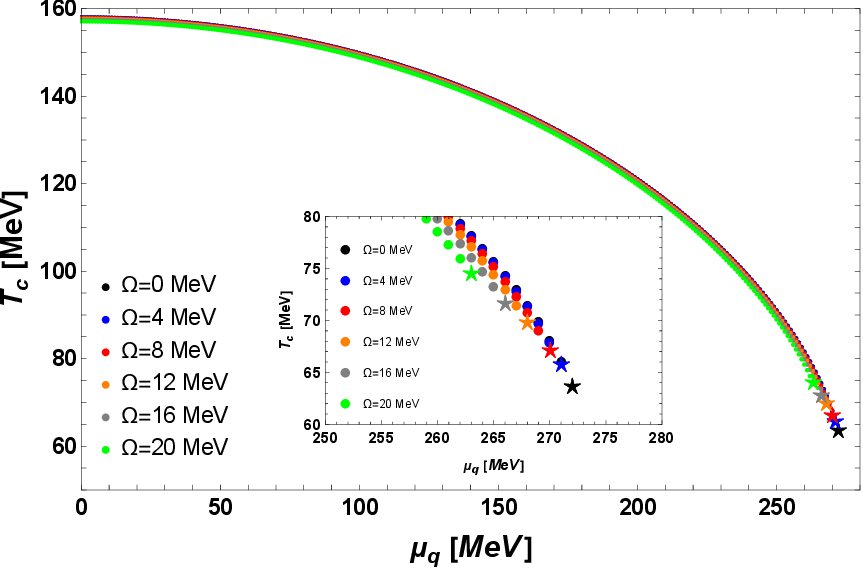}
    \caption{Phase diagram, in the temperature quark chemical potential plane, obtained from the high temperature expression for the effective potential for $\lambda=1.4$, $g=0.88$, $a=148.7$ MeV and $\mu=500$ MeV. For six different values of the angular velocity. All the dots represent a second-order phase transition and the CEP is represented by an star in each of the six different cases.}
    \label{fig3}
\end{figure}

\begin{figure}[t]
    \centering
    \includegraphics[scale=0.58]{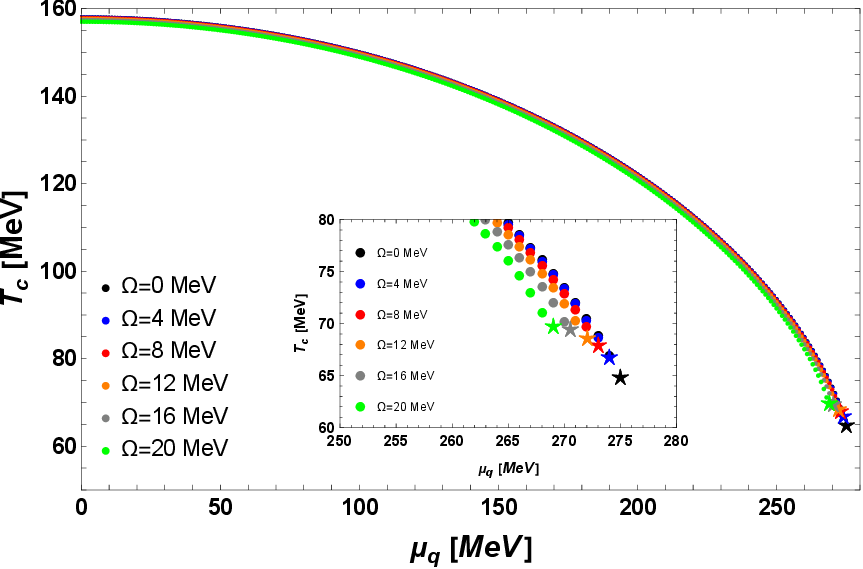}
    \caption{Phase diagram, in the temperature quark chemical potential plane, obtained from the high temperature expression for the effective potential for $\lambda=1.4$, $g=0.836$, $a=143.2$ MeV and $\mu=500$ MeV. For six different values of the angular velocity. All the dots represent a second-order phase transition and the CEP is represented by an star in each of the six different cases.}
    \label{fig4}
\end{figure}

In Fig.~\ref{fig3}, we find the effective QCD phase diagram, using the values $\lambda=1.4$, $g=0.88$ and $a=148.7$ MeV. We plot six different curves, each of them for a given value of angular velocity, we choose $\Omega=0, \ 4, \ 8, \ 12, \ 16, \ 20$ MeV. What is most striking are two effects. On the one hand, we notice that the critical temperature decreases as the angular velocity increases. At the same time, we also observe that the CEP moves towards lower values of the quark chemical potential and towards higher values of temperature, when the angular velocity increases. It tells us that the angular velocity is able to not only change the conditions where the phase transition occurs but also can modify the nature of the phase transition. Since, the result obtained is one the central goals of this work, we repeat the analysis with other set of parameters. It is a probe that can shows the parameters' independent result. Fig.~(\ref{fig4}) shows a phase diagram with $\lambda=1.4$, $g=0.836$ and $a=143.2$ MeV, for the same set of $\Omega$'s values, and we see the same behavior of $T_c$ and displacement of the CEP as a function of the angular velocity.

\section{Baryon Number Fluctuation \label{sec4}}

\begin{figure}[t]
    \centering
    \includegraphics[scale=0.58]{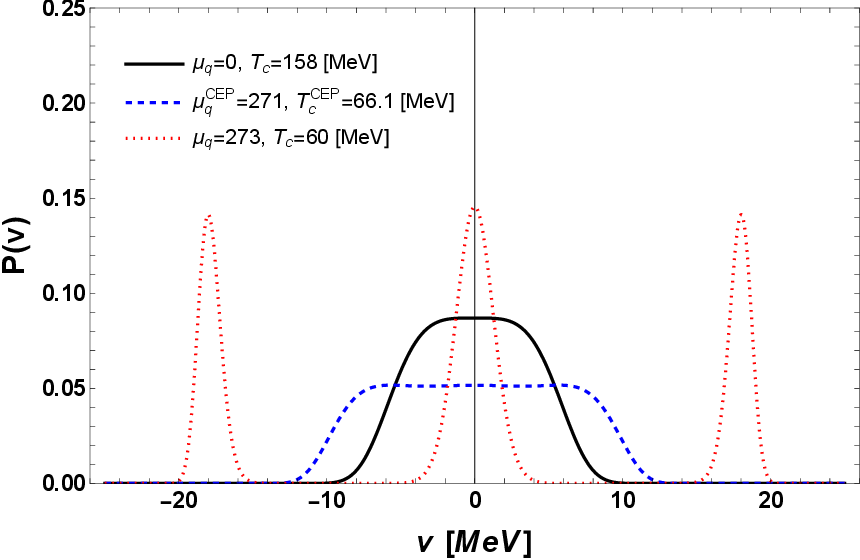}
    \caption{Normalized probability distribution as a function of $|v|$ for three different points along the phase transition for a fixed angular velocity $\Omega=16$ MeV. $\mu_q=0$ MeV and $T_c=158$ MeV corresponds to second-order phase transition, $\mu_q^{\text{CEP}}=266$ MeV and $T_c^{\text{CEP}}=71.7$ MeV corresponds to the CEP, and $\mu_q=269$ MeV and $T_c=65.5$ MeV is where we find a first-order phase transition.}
    \label{fig5}
\end{figure}

In order to complement and use the results obtained from the analysis of the effective potential using the LSMq in the previous section. We proceed to compute the fluctuation in the baryon number. To achieve this,  we start with the expression for the probability distribution 
\begin{equation}
    \mathcal{P}(v)=e^{-\mathcal{V}V^{eff}(v)/T},
    \label{distribution}
\end{equation}
where $v$ is the order parameter and $\mathcal{V}$ is the volume of the system. The properties of Eq.~(\ref{distribution}) can be obtained studying the behavior from the statistical moments or cumulants. However, for this work we are focused just on the fourth moment $\kappa \sigma^2=c_4/c_2$. We examine the dependence of $\kappa \sigma^2$ on the collision energy in heavy-ion reactions $\sqrt{s_{NN}}$, where here $\sigma^2$ is the variance. Before to do the analysis of the fourth moment, the probability distribution itself provides important insights of the phase transition order. In Fig.~(\ref{fig5}), we plot the normalized probability distribution as a function of $|v|$, for three different points along the phase transition curve depicted in Fig.~(\ref{fig3}), with a fixed value of angular velocity, $\Omega=16$ MeV. The values of $\mu_q$ and $T$ for these three points chosen are $\mu_q=0$ MeV and $T_c=158$ MeV, $\mu_q=266$ MeV and $T_c=71.7$ MeV, and $\mu_q=269$ MeV and $T_c=65.5$ MeV, which correspond to the case of a second-order phase transition, the critical end point and a first-order phase transition, respectively. The shape of $\mathcal{P}$ for the second-order phase transition is Gaussian-like. The CEP provides the widest probability distribution that can be achieved, and the first-order phase transition shows a narrow peak at the center together with other two equidistant peaks, with the same high all of them, and it is a signal of a degenerated minima.  

We now proceed to do our last analysis. Hence, we show from Figs.~(\ref{Kurtosis}) and~(\ref{fig8}), the behavior of the fourth moment normalized to the same quantity computed for $\mu_B=0$ and $T=T_c$, as functions of the collision energy in heavy-ion reactions, where we remember the relation $\mu_q=\mu_B/3$. For this purpose, we resort to the relation between the chemical freeze-out value of $\mu_B$ and the collision energy, $\sqrt{s_{NN}}$, given by~\cite{Cleymans:2005xv,Randrup:2006nr}
\begin{equation}
    \mu_B(\sqrt{s_{NN}})=\frac{d}{1+e\sqrt{s_{NN}}},
    \label{Cleynmanseq}
\end{equation}
where $d=1.308$ GeV and $e=0.273$ GeV$^{-1}$. Figures~\ref{Kurtosis} is $\kappa \sigma^2$, normalized to the same fourth moment for $\mu_B=0$ and $T=T_c$, for the upper panel three different values of angular velocity $\Omega=0, \ 8$ and $16$ MeV, and for the lower panel there are other three different values of angular velocity $\Omega=4, \ 12$ and $20$ MeV. It is computed with the set of parameters $\lambda=$, $g=$ and $a=$ which are the same that we used in Fig.~\ref{fig3}. The value of $\sqrt{s_{NN}}$ for each plot that corresponds to the CEP location is represented by vertical lines.

Figure~\ref{fig8} shows $\kappa \sigma^2$, normalized to the same fourth moment for $\mu_B=0$ and $T=T_c$, as a function of $\sqrt{s_{NN}}$, with the same set of parameters used in Figs.~\ref{Kurtosis}, for three different values of the reaction volume, $\mathcal{V}= 50^3, \ 100^3$ and $150^3 \ \text{fm}^3$. One more time, the value of $\sqrt{s_{NN}}$ for each plot that corresponds to the CEP location is represented by a vertical line.

\section{Results \label{sec5}}

\begin{figure}[t]
        \centering
    \includegraphics[scale=0.58]{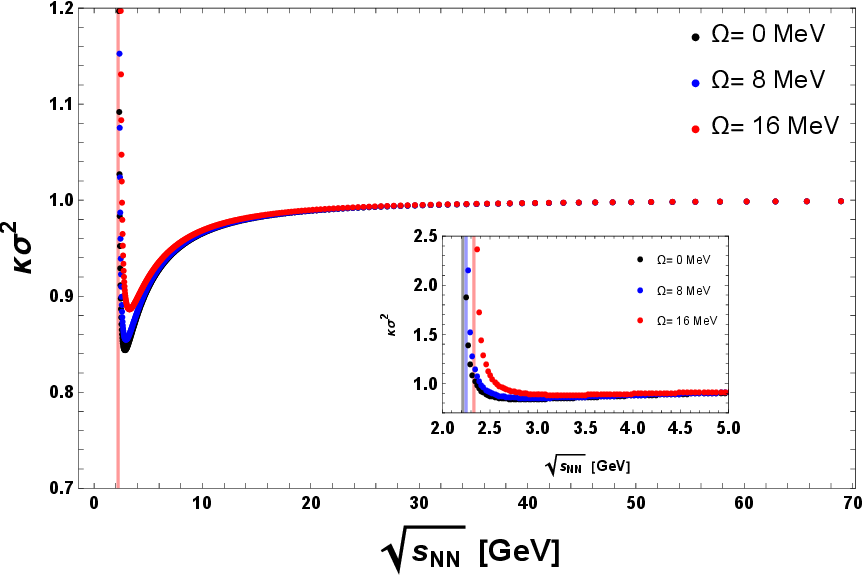}
    \\[\bigskipamount]
        \centering
    \includegraphics[scale=0.58]{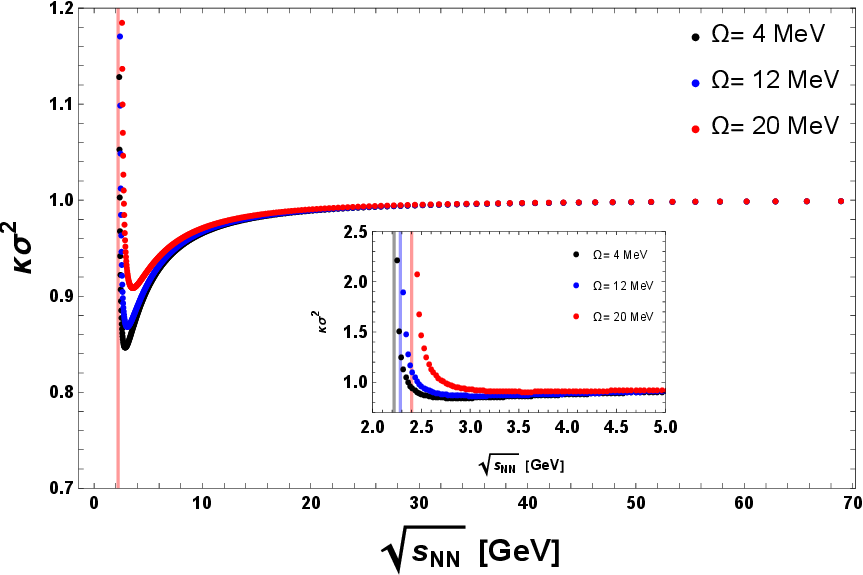}
    \caption{Fourth moment $\kappa \sigma^2$ normalized to the same quantity computed for $\mu_B=0$ and $T=T_c$, for different values of $\Omega$ as a function of the collision energy in heavy-ion reactions $\sqrt{s_{NN}}$, using its relation with $\mu_B$ given by Eq.~(\ref{Cleynmanseq}). Both panels are computed with the same set of parameters used in Fig.~\ref{fig3}. However, the upper panel shows three different values of angular velocity $\Omega=0, \ 8$ and $16$ MeV and the lower panel shows other set of three angular velocities $\Omega=4, \ 12$ and $20$ MeV. For values of $\sqrt{s_{NN}}$ between 2 
 and 2.5 GeV, where a vertical lines are plotted, corresponds to the region where the CEP is located.}
    \label{Kurtosis}
\end{figure}

The analysis of the phase transition, related with the chiral symmetry restoration at finite temperature, quark chemical potential and angular velocity, carries out in this work can be divided in two main parts. The first one  are the results shown in Figs.~\ref{fig3} and~\ref{fig4}. They are effective QCD phase diagrams which shows how the critical temperature $T_c$ changes for a wide region of quark chemical values $\mu_q$. It is obtained from the effective potential computed in Sec.~\ref{sec3}. In each phase diagram, we plot the phase transition curve from $\mu_q=0$ up to the critical end point, for six different values of angular velocity $\Omega$. Beyond the CEP, we find only first-order phase transition, in each of the six cases. It is noteworthy that as $\Omega$ increases, the phase transition occurs at progressively lower values, indicating that angular velocity catalyzes symmetry restoration. Also, in the effective phase diagrams, we are able to observe how the CEP moves as a function of $\Omega$, if the angular velocity increases the CEP moves for lower values of $\mu_q$ and larger values of $T$. It means that the onset of a first-order phase transition can be found more quickly if we increase the angular velocity. This is a result which can be contrasted with Ref.~\cite{Chen:2023cjt}, where their result shows that the behavior of the transition is in agreement with our result, but the CEP slightly moves to lower temperature and lower density. In Ref.~\cite{Singha:2024tpo} the authors reported the same behavior of the transition lines as a function of $\Omega$ at zero quark chemical potential. Moreover, the behavior of the transition curves and the changes of the CEP as a function of $\Omega$ is in contradiction to the results reported in Ref.~\cite{Chen:2024jet}. Also, the result we find exhibits the same kind of behavior found in the QCD phase diagram when a uniform and constant magnetic field is present~\cite{Ayala:2021nhx}. The second part of the analysis focuses on baryon number fluctuations. In Figs.~\ref{Kurtosis}, we show the normalized fourth moment $\kappa\sigma^2=c_4/c_2$ referenced to its value at $\mu_B=0$ and $T=T_c$ as a function of the collision energy in heavy-ion reactions $\sqrt{s_{NN}}$. From these plots, we observe an almost constant behavior for $\sqrt{s_{NN}}$ between $20$ GeV and $70$ GeV, but for lower values of the collision energy we see that the CEP position is heralded not by the dip of $\kappa \sigma^2$ but for its strong rise as the energy that corresponds to the CEP is approached. A similar result without the vortical effects has been found in Refs.~\cite{HADES:2020wpc,Mroczek:2020rpm,Braun-Munzinger:2020jbk,Vovchenko:2021kxx,STAR:2022etb,Ayala:2021tkm}. Beyond the agreement of our results with others reported in the literature, we want to highlight a relevant behavior in the Figs.~\ref{Kurtosis}. When the value of the angular velocity increases, we see that the place where the CEP appears changes, such that at higher values of $\Omega$, the CEP moves to larger values of collision energy. The latter generates a greater agreement between what was obtained in this work and the experimental results~\cite{HADES:2020wpc,STAR:2022etb} reported so far.

\section{Conclusions \label{sec6}}

In this work, we have studied the impact of vortical effects on the dynamics of strongly interacting matter, using the LSMq. This model has proven to be an effective tool to analyze the restoration of chiral symmetry under extreme conditions, such as high temperatures, densities and the presence of vorticity. Throughout the analysis, we have gone beyond the mean-field approximation by considering contributions from the ring diagrams, which has allowed us to incorporate screening effects in the medium. One of the main results of this work is the construction of an effective QCD phase diagram, where we show how the critical temperature $T_c$ varies with the quark chemical potential $\mu_q$ and the angular velocity $\Omega$. We find that as $\Omega$ increases, the critical temperature decreases, indicating that vorticity accelerates the restoration of chiral symmetry. Furthermore, we have observed a significant shift of the CEP in the phase diagram. With increasing $\Omega$, the CEP shifts towards lower $\mu_q$ values and higher $T_c$, suggesting that the angular velocity not only alters the conditions under which the phase transition occurs, but also the nature of the transition, favoring the appearance of first-order transitions at lower $\mu_q$. Another significant result of this work is the analysis of baryon number fluctuations via the normalized fourth moment $\kappa \sigma^2 = c_4 / c_2$. We have shown that the fourth moment exhibits an almost constant behavior for large collision energies $\sqrt{s_{NN}}$. However, as the energy approaches the CEP position, the fourth moment exhibits a sharp increase, suggesting that the CEP location can be identified by this abrupt rise. This behavior is also influenced by vorticity, as higher values of $\Omega$ shift the CEP to higher collision energies. This finding is particularly relevant and could be contrasted with experimental results obtained in heavy ion collisions. These findings open new avenues for exploring the effects of vorticity in future experimental investigations of heavy ion collisions, particularly in projects such as NICA, FAIR and the Beam Energy Scan program at RHIC.

\begin{figure}
    \centering
    \includegraphics[scale=0.58]{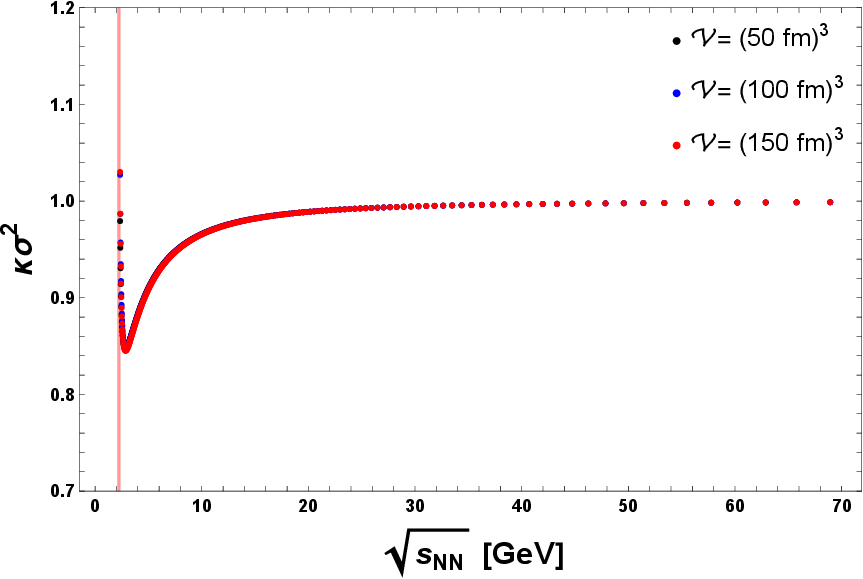}
    \caption{Fourth moment $\kappa \sigma^2$ normalized to the same quantity computed for $\mu_B=0$ and $T=T_c$, for three different values of the volume $\mathcal{V}$ as a function of the collision energy in heavy-ion reactions $\sqrt{s_{NN}}$, using its relation with $\mu_B$ given by Eq.~(\ref{Cleynmanseq}). It is computed with $\Omega=0$ MeV and the same set of parameters used in Fig.~\ref{fig3}. From this plot, we observe that the effect of the volume is negligible.}
    \label{fig8}
\end{figure}

\begin{acknowledgments}
Support for this work was received in part by Consejo Nacional de Humanidades, Ciencia y Tecnolog\'ia Grant No. CF-2023-G-433. L. A. H. acknowledges support from DCBI UAM-I PEAPDI 2024, and DAI UAM PIPAIR 2024 projects under Grant No. TR2024-800-00744. R.Z. acknowledge support from ANID/CONICYT FONDECYT Regular (Chile) under Grant No. 1241436.
\end{acknowledgments}

\appendix

\section{Matter piece of 1-loop fermion contribution in the effective potential\label{Appe1}}
We start with the 1-loop fermion contribution
\begin{align}
    V_{\text{f}}^1=-T \sum_n \int dm_f^2 &\int \frac{d^3k}{(2\pi)^3} \Big [ \frac{1}{(\omega_n-i\mu_1)^2+\vec{k}^2+m_f^2} \nonumber \\
    &+ \frac{1}{(\omega_n-i\mu_2)^2+\vec{k}^2+m_f^2} \Big], 
\end{align}
performing the sum over the Matsubara modes, we have
\begin{align}
    V_{\text{f}}^1&=-\int dm_f^2 \int \frac{d^3k}{(2\pi)^3}\frac{1}{2E}\nonumber \\
    &\times \bigg[ \Big(1-n_f(E+\mu_1)-n_f(E-\mu_1) \Big)\nonumber \\
    &+\Big(1-n_f(E+\mu_2)-n_f(E-\mu_2) \Big)\bigg].   
\end{align}
We continue the calculation only working on the matter term. Thus, we have
\begin{align}
    V^1_{\text{f,mat}}&=-\int dm_f^2 \int \frac{d^3k}{(2\pi)^3}\frac{1}{2E}\nonumber \\
    &\times \bigg[ \Big(-n_f(E+\mu_1)-n_f(E-\mu_1) \Big)\nonumber \\
    &+\Big(-n_f(E+\mu_2)-n_f(E-\mu_2) \Big)\bigg],   
\end{align}
we integrate over $m_f^2$ and obtain
\begin{eqnarray}
V^1_{\text{f,mat}}&=&-T \int \frac{d^3k}{(2\pi)^3} \bigg[ \ln(1+e^{-(E-\mu_1)/T})\nonumber \\
&+&\ln(1+e^{-(E+\mu_1)/T})+\ln(1+e^{-(E-\mu_2)/T}) \nonumber \\
&+&\ln(1+e^{-(E+\mu_2)/T})\bigg].
\label{V1fmatafterMatSum}
\end{eqnarray}
The Eq.~(\ref{V1fmatafterMatSum}) is the same as Eq. (A.137) of \cite{kapusta_gale_2006} with $(2s+1)=-1$. Therefore, it becomes
\begin{eqnarray}
V^1_{\text{f,mat}}&=&-\frac{m_f^2T^2}{2\pi^2}\sum_{l=1}^{\infty}\frac{(-1)^{l+1}}{l^2}K_2(lm_f/T) \nonumber \\
&\times&\left( e^{l \mu_1/T} + e^{-l \mu_1/T}+e^{l \mu_2/T}+e^{-l \mu_2/T} \right). \label{A6}
\end{eqnarray}
Finally, we take the high temperature limit in Eq.~(\ref{A6}), compute the sum over $l$ and obtain
\begin{align}
    V^1_{\text{f,mat}}=&\frac{m_f^4}{16\pi^2}\bigg [ \ln \Big(\frac{m_f^2}{\pi^2 T^2} \Big) + 2\gamma_E -\frac{3}{2} \bigg]-\frac{7 T^4 \pi^2}{180}\nonumber \\
&-\frac{T^2}{12}\Bigg( \Big( \mu_q+\frac{\Omega}{2} \Big)^2+\Big( \mu_q-\frac{\Omega}{2} \Big)^2 \Bigg)\nonumber \\
&-\frac{T^2m_f^2}{4\pi^2}\Bigg ( \text{Li}_2\left(-e^{\frac{\mu +\frac{\Omega }{2}}{T}}\right)+\text{Li}_2\left(-e^{\frac{\mu -\frac{\Omega }{2}}{T}}\right)\nonumber \\
&+\text{Li}_2\left(-e^{-\frac{\mu +\frac{\Omega }{2}}{T}}\right)+\text{Li}_2\left(-e^{-\frac{\mu -\frac{\Omega }{2}}{T}}\right) \Bigg ) \nonumber \\
&-\frac{\left(\mu +\frac{\Omega }{2}\right)^4+\left(\mu -\frac{\Omega }{2}\right)^4}{24 \pi ^2}.
\end{align}

\section{Boson’s self-energy in the hight temperature limit\label{appe2}}
The boson's self-energy can be splitted in two term, the bosonic and fermionic contributions
\begin{equation}
\Pi=\Pi_b + \Pi_f,
\end{equation}
where the bosonic term $\Pi_b$ was reported in \cite{Ayala:2017ucc} and the final expression is
\begin{equation}
\Pi_b =\frac{\lambda T^2}{2}.
\label{PibhighT}
\end{equation}
For the femionic term, we use the relation
\begin{equation}
\Pi_f= N_f N_c 2 g^2 \frac{d V^1_{\text{f,mat}}}{d m_f^2},
\end{equation}
which is valid only assuming that $T$ is the largest energy scale.
If we perform the derivative and keep only the leading term in $T$, we obtain
\begin{eqnarray}
\Pi_f&=& -\frac{N_f N_c T^2 g^2}{2 \pi^2} \Bigg ( \text{Li}_2\left(-e^{\frac{\mu +\frac{\Omega }{2}}{T}}\right)+\text{Li}_2\left(-e^{\frac{\mu -\frac{\Omega }{2}}{T}}\right)\nonumber \\
&+&\text{Li}_2\left(-e^{-\frac{\mu +\frac{\Omega }{2}}{T}}\right)+\text{Li}_2\left(-e^{-\frac{\mu -\frac{\Omega }{2}}{T}}\right) \Bigg ).
\label{PifhighT}
\end{eqnarray}
As a final step, we put together the Eqs.~(\ref{PibhighT}) and~(\ref{PifhighT}) and the boson's self-energy becomes
\begin{eqnarray}
\Pi&=& \frac{\lambda T^2}{2} -\frac{N_f N_c T^2 g^2}{2 \pi^2} \Bigg ( \text{Li}_2\left(-e^{\frac{\mu +\frac{\Omega }{2}}{T}}\right)+\text{Li}_2\left(-e^{\frac{\mu -\frac{\Omega }{2}}{T}}\right)\nonumber \\
&+&\text{Li}_2\left(-e^{-\frac{\mu +\frac{\Omega }{2}}{T}}\right)+\text{Li}_2\left(-e^{-\frac{\mu -\frac{\Omega }{2}}{T}}\right) \Bigg ).
\end{eqnarray}

\bibliography{mybibliography}

\end{document}